# Orchestrating Metadata Enhancement Services: Introducing Lenny

Jon Phipps
National Science Digital Library
(NSDL) Cornell University
301 College Ave.
Ithaca, NY 14850
+1 607 255 8510

jphipps@cs.cornell.edu

Diane I. Hillmann
National Science Digital Library
(NSDL) Cornell University
301 College Ave.
Ithaca, NY 14850
+1 607 255 5691

dih1@cornell.edu

Gordon Paynter
The INFOMINE Project
Science Library,
University of California
Riverside, CA 92517-5900
+1 951 827 2279

paynter@library.ucr.edu

## ABSTRACT
Harvested metadata often suffers from uneven quality to the point that utility is compromised. Although some aggregators have developed methods for evaluating and repairing specific metadata problems, it has been unclear how these methods might be scaled into services that can be used within an automated production environment. The National Science Digital Library (NSDL), as part of its work with INFOMINE, has developed a model of service interaction that enables loosely-coupled third party services to provide metadata enhancements to a central repository, with interactions orchestrated by a centralized software application.

## Categories and Subject Descriptors
H.3.7 [Information Storage and Retrieval]: Digital Libraries—*collection, standards, system issues.*

## 1. Introduction
The problem of metadata quality has been with us ever since the first librarians wandered out from their warm traditional environment into the maelstrom of the world beyond MARC. In the MARC world, when quality was at issue, the wagons circled and the errant practitioners were advised to bring their metadata back in line with the norm, so that the carefully organized distribution and reuse mechanisms continued to function predictably.

In the metadata world, there is precious little existing consistent practice at the data provider end of the equation, and no sanctions worth mentioning for the crime of bad metadata. Thus it has behooved aggregators to find other methods to ensure that records harvested from diverse providers are of sufficient quality to meet user needs for discovery and use of resources.

Work in this area has been building since the first discussions of the problem. Dushay and Hillmann [1] described a number of significant problems with metadata harvested into the nascent NSDL, and others have discussed similar problems in other aggregated environments [2], [3], [4].

In this paper we describe collaborative work between the NSDL and INFOMINE which has resulted in a model of service interactions, optimizing reliability with minimal human intervention. We believe this model serves the developing world of metadata aggregation and re-use well, providing an extensible method of improving metadata as well as a model for developing metadata services in an interoperable environment where they can be used by a variety of digital libraries.

## 2. Background
The NSDL is a program of the National Science Foundation, engaged in building library collections and services for all aspects of science education [5]. Now in its fourth year of operation, the NSDL is building upon a technical foundation established early in its development (and described more fully in [6], [7]). As part of its mission, the NSDL gathers and updates increasing amounts of metadata pertaining to resources in the fields of science, technology, engineering and mathematics, aiming for high quality, large quantity and low cost approaches.

The original organizing principle behind the NSDL repository is the archive-inspired "collection/item" limited hierarchy, which views the library as a set of collections, each consisting of many items, and assumes that each item in a collection is a resource described by a complete "item-level" metadata record. In the NSDL, a single resource, identified by a URI, may be contained in many collections and described by many different item-level records. These item-level descriptions are maintained in the metadata repository as entities separate from the resource they're describing and will be combined on output to form a single aggregated, hopefully comprehensive, description of that resource [8].

## 2.1 Rethinking the Framework

The collection/item model grew increasingly strained as the poor quality of the metadata made available by some collection-based metadata providers, and the absence of metadata from many others, affected the quality of the NSDL repository as a whole. Even those that provided metadata of good quality too often saw resources through the lens of a particular specialty or domain.

An early service of the NSDL was to help collection holders to provide high-quality metadata via the Open Archives Initiative Protocol for Metadata Harvesting (OAI-PMH). Although it was not a surprise that the initial acceptance of OAI-PMH was limited to the larger, better-funded and more technically savvy collections, early ideas concerning the development of mechanisms of collection-based metadata provision were naïve at best. When it became clear that NSDL staff could not single-handedly assist all relevant collections to become OAI-PMH data providers, and that some collections were unable or unwilling to make the necessary efforts, alternative strategies were developed.

One shift in thinking came with the recognition that even collections unable to provide metadata were still providing a service—a "selection service"—by choosing from among the available resources in their topic area those which fit their criteria for inclusion. Any collection, then, could be seen as a potential provider of two distinct services: selection, and metadata creation. If a collection could not or would not provide metadata, another service might be able to provide it.

This new notion of "service" complements NSDL's evolving notions of metadata augmentation and recombination. In particular, exposure of detailed information about the source of metadata recognizes this split between selection services and metadata providers. The former are identified on the public NSDL interface by selection service "brands," while the latter can be identified in our metadata dissemination format designed for this purpose, which we call "augmented" metadata.

The new services providing metadata separately from selection could be developed either by NSDL or by others, but either way they needed to be seen by automated systems as separate, independent entities, providing data in standard formats with standard interfaces. Clearly, with this basic scenario, the challenge is to orchestrate these interactions in ways that emphasize automated solutions.

## 2.2 An Orchestra of Services

Using the term "orchestra" to describe the array of services we envision emphasizes two important characteristics of this approach. First, the services we describe below are external to the harvesting task; each a separate player, specialized to perform particular operations, who can be called upon to provide specific outputs. Second, these complex parts are coordinated by a central intelligence—a "conductor" who is not a player, but whose contribution is vital to the performance as a whole. Each of the services has their own roles, characteristics and relationships with other players—they resemble a group of musicians waiting for their cues. We describe these services below.

### 2.2.1 Primary metadata generation services

The current NSDL architecture expands the simple collection/item association by ascribing to the collection owner the role of "selector." This allows the separation of that function, essentially one of collection development (in traditional library terms), from that of metadata provider, thus addressing the reality that metadata can not always be provided by the same entity that selected the resources for inclusion into the collection.

In the basic scenario currently used in building the NSDL, a website or aggregation of materials is selected by a subject expert, and incorporated into the NSDL repository via a semi-automated process of creating a descriptive collection record. If an OAI-PMH server is available for the collection, it is designated as the primary metadata provider. If it is not, an automated crawl of the site can be invoked from an external service provider (such as INFOMINE), and the automated generation service becomes the primary provider of metadata for the collection, even though the items in the collection were selected by another entity. This strategy works best with openly accessible sites containing adequate text.

### 2.2.2 Metadata augmentation services

If metadata is provided by the collection creator but has gaps in important areas (e.g., missing data elements), an automated metadata augmentation service can provide an additional pass at metadata provision. In this case the augmentation service harvests the item records from the NSDL, crawls the URLs contained in the item metadata records, then exposes new metadata to the repository for harvest. Metadata augmentation services are particularly useful for enriching records lacking specific elements, such as subject headings or media types, that are part of the metadata element set most useful to the NSDL.

### 2.2.3 Transformation (safe and collection-specific) services

Safe transformations are those that improve the consistency of metadata and can be accomplished on any metadata without danger of degrading the original [9]. For example, it is always safe to alter the case of a descriptor from a controlled vocabulary like the Library of Congress Subject Headings (LCSH) when it occurs in a non-standard format.

Collection-specific transformations require an evaluation of an individual provider's current practices, and operations made in that context cannot necessarily be generalized to other collections or even to the same collection over time. In the NSDL's augmentation scenarios, the data is not replaced when transformed via safe or collection-specific services, but it is added to the full, recombined metadata as additional information. For instance, if a data provider included a format value of "HTML," the collection-specific transformation might add the statement

*<dc:format xsi-type="dct:IMT">text/html</dc:format>*

to all item records with that value.

In combination these two services provide cost-effective and configurable methods to improve the consistency and usefulness of metadata for downstream users of NSDL data. An added bonus is that these transformation services can boost Simple oai_dc metadata to more robust nsdl_dc (the NSDL's qualified DC schema) by recognizing values from recommended vocabularies and ascribing the appropriate vocabulary encoding scheme to the statements.

### 2.2.4 Equivalence services

Central to the notion of recombined metadata is the ability to associate metadata from multiple providers to provide a fuller description of a resource. A prerequisite is the ability to discern when two item-level records describe the same resource. In most cases, a simple URI equivalence is sufficient, but experience has indicated that the simple approach is insufficient where the URI or other standard identifier varies to some extent, though the resource itself is the same. An equivalence service allows the association between resources identified by variant identifiers to be made in a reliable manner, enabling the recombination of descriptive statements from a variety of metadata records into a fuller description.

Simple equivalence services rely on associating URLs and common URL variants to assert equivalence. Down the line, more sophisticated services might depend on newly emerging and maturing technologies to extend the ability to associate equivalent resources despite differences in format or location.

### 2.2.5 Crosswalking (schema and vocabulary)

Crosswalking services are a type of metadata augmentation service that generate new fielded metadata values based on a crosswalk from one schema or vocabulary to another. A service providing crosswalks between metadata schemas (also called element sets), might have access to a variety of standard crosswalks. Godby, Young and Childress [10] describe how such a metadata schema crosswalking service might operate within this orchestra of services.

Value-based crosswalking services, operating on controlled and uncontrolled vocabulary value strings associated with specific elements, could be organized in a similar manner, as similar components would be needed to support such services in an automated environment. Both kinds of crosswalking services improve the ability of metadata to be reused in a variety of knowledge domains, expressing attributes of resources in terms familiar to particular groups of users and enabling more effective filtering of search results for them.

### 2.2.6 Archiving/Persistence services

Archiving and persistence services ensure that the content to which metadata refers is always available to users. In the NSDL context, these services rely on Web crawling to harvest content, and can potentially provide alternative URLs that link to cached copies of the content when the originals no longer lead to the described resource. These alternative URLs could reside in NSDL augmented metadata, since the attribution of source allows distinctions to be made between links to original and cached content.

### 2.2.7 Annotation services

An Annotation service is a special form of metadata augmentation service that provides third-party resource reviews. Oftentimes, the collections that gather resources within a topic area also allow their users to review, rate or contribute comments about their resources, either formally or informally. These annotations about the resources give a different perspective on possible uses, and may also, when formally developed, apply a specific set of criteria to the evaluation of the resource. Such information is normally stored and maintained by the originating portal or collection, but if portal owners make metadata for these associated resources available for NSDL portals, they can be served to users with search results.

### 2.2.8 Metadata improvement and rating services

The process of building up augmented and recombined metadata necessarily carries the implication that some metadata statements are "better" than others. Better may mean more reliable, more useful, or simply a preponderance of a particular value among a number of available statements. As an example, if a dozen statements about the format of an item exist, and ten say it's HTML, the assertion that the item is really HTML might be given more credence. Collectively, the "safe" and collection-specific transformations will operate as improvement services, providing higher quality statements that operate more predictably in an aggregated environment.

Given that recombined metadata contributed and "improved" by a variety of services might carry a great number of individual statements, some of them possibly contradictory or confusing, there are several ways that a downstream user might react to the improved metadata made available by the NSDL. They might take all the statements and treat them as equally useful, hoping their users will sort things out eventually. Or, they might evaluate the statements and sources offered and figure out for themselves which were reliable and useful, based on source information provided with the data. A third option is that the NSDL provide its selection of statements as an additional service, providing a "Gold" aggregation of statements that provide the most reliable and useful information for most downstream users.

## 2.3 An example

The following example illustrates how NSDL employs several different services to build an aggregated set of item-level statements for a new collection.

> *Primary Metadata Generation*: NSDL creates a new collection-level record for the Whatsis Collection, and schedules monthly harvests from its OAI-PMH server. At each harvest, NSDL receives a set of item records from the Whatsis Collection. The NSDL repository can now redistribute the collection's item-level records through its OAI-PMH server using the standard oai_dc and nsdl_dc formats.

> *Safe Transformation*: As it happens, the DCMIType value of "Interactive Resource" is consistently misspelled by the Whatsis Collection provider in the dc:type element. Since this is a common misspelling routinely looked for, a second version of the dc:type element with the correctly spelled value is provided by the NSDL "safe" transform service, and the encoding scheme of dct:DCMIType is added [11]. A metadata correction notification message is sent to the data provider. The nsdl_dc_plus record will include both versions of dc:type; nsdl_dc_gold will only show the correctly spelled one, with its indicated encoding scheme.

> *Collection-specific transformation*: Many Whatsis item records contain the dc:publisher value of "The University," carried over from a MARC record 260 $b. Since all the records in the collection that use this value also include a DC creator value of Pennsylvania State University, an additional dc:publisher element with the fuller version of the value is provided by the NSDL collection-specific transformation service, and a metadata correction notification message is sent to the data provider. The Repository will serve an OAI-PMH format called nsdl_dc_plus which will include both versions of dc:publisher; the served nsdl_dc_gold format will only show the more correct version supplied by the transformation service.

*Metadata Augmentation*: During the evaluation of the first harvest from this provider, the NSDL editor notes that the items lack subject information. To remedy this, the INFOMINE Metadata Enrichment Service is invoked, and used to automatically generate an additional set of item-level metadata records matched to the resources in the collection, but containing additional subject information, including LCSH and keywords. These added subject elements will be included in both nsdl_dc_plus and nsdl_dc_gold records provided by the NSDL OAI-PMH server.

## 2.4 Introducing Lenny

"Lenny" is the portion of the NSDL's Collection Registration Service (CRS) that "conducts" the interactions between the NSDL's harvest and ingest processes and the services that provide data to those processes. In the example above, Lenny is responsible for scheduling the OAI-PMH harvests from services, and invoking the transformation services and metadata augmentation services that contribute to the item-level records.

To accomplish these tasks, the CRS provides a set of interfaces that allow NSDL's human editors to describe each collection and service to Lenny. The process begins as NSDL subject experts recommend resources for inclusion in the library. NSDL editors review these submissions, identify resources that contain useful item-level material, and designate them as collections. The editor must then identify a service providing item-level metadata for each collection, provide a location from which the metadata may be harvested, and schedule the flow of metadata into the NSDL Repository.

Lenny provides facilities for scheduling full and incremental OAI-PMH harvests, and for selecting specific OAI-PMH sets and metadata formats for each harvest to support all the services described above. To prevent cacophony, the interactions of these services between Lenny, the NSDL repository, and each other need to be carefully managed, with careful attention to order, timing and response.

Many services, once invoked, are free to harvest their primary source metadata from the NSDL Repository and provide their results asynchronously as it becomes available. Lenny has ensured that the primary metadata that is their only prerequisite has already been successfully stored in the NSDL Repository before they're called on to play. Because the Transformation and Augmentation services rely on already existing metadata (either existing statements or URLs) these services must wait until some primary metadata is provided for the resources aggregated in a collection.

## 2.5 Service Interactions, Web Services and Lenny

Lenny's service orchestration model uses OAI-PMH because it has been specifically designed to efficiently transport blocks of metadata in the Representational State Transfer (REST) [12] environment of the internet, forming a good service model for providing metadata to the NSDL Repository. The services being conducted by Lenny are loosely coupled by the very nature of the OAI-PMH, but tightly coupled through the need for service interactions to be frequently and properly sequenced.

Lenny supports two methods of service coordination, passive or interactive, and each has its place in the overall orchestration of data flow. Passive services repeatedly harvest metadata on a fixed schedule determined by the service. The service then performs some operation, and makes its results available for OAI-PMH harvest by Lenny. Lenny may request a harvest of the results on a similarly fixed schedule or on demand. These services behave much like soloists on a riff – they just start playing, and Lenny and the rest of the band wait a predetermined time for them to finish before starting to play again. For example, the NSDL Archive Service harvests newly updated items in the Repository on its own schedule without interaction or scheduling by Lenny. Lenny can then harvest information on links for cached content for use within the NSDL public portal.

Interactive services require a greater degree of coordination. Lenny orchestrates interactive services through an event-driven messaging interface. Service administrators provide an interface that Lenny can use to trigger the service's functions. Lenny provides an interface that allows services to notify Lenny of failure or completion of the request, provide the location of logs (if available) and, of course, the OAI-PMH request that will retrieve the desired data. These interfaces can be powered by CGI scripts [13], XML-RPC [14] or SOAP [15], but it's important that the services stick to whatever interaction protocol was set up when the service was registered. At present, the Metadata Provider Services working with the NSDL operate as interactive services, A few of these services must be run in sequence, but most can be run independently of each other.

When Lenny receives notification that an interactive service has completed its requested process successfully, a harvest is invoked. If the service notifies Lenny of failure, a notification containing available details about the failure is sent to NSDL editors and to the service administrator. The results of each of Lenny's requests, including request parameters, are maintained in an internal log. Services can be invoked on a repeating schedule, and each scheduled invocation can have its own set of parameters. For instance a complete harvest of a metadata provider can be run every 6 months and an incremental harvest of that same provider every day.

Because resources on the web are always subject to change, part of the job of metadata provider services is to help the NSDL keep its metadata fresh. The incremental harvest feature of OAI-PMH supports provision of fresh updates of metadata from most service providers. For instance, once the INFOMINE Expert-Guided Crawler has created an initial data set, Lenny will then request an incremental re-crawl of the site on a pre-determined schedule. The Expert-Guided Crawler will perform a complete re-crawl of the site, updating the metadata only for pages that have changed, making the freshened metadata records available to the NSDL Repository as an OAI-PMH incremental harvest.

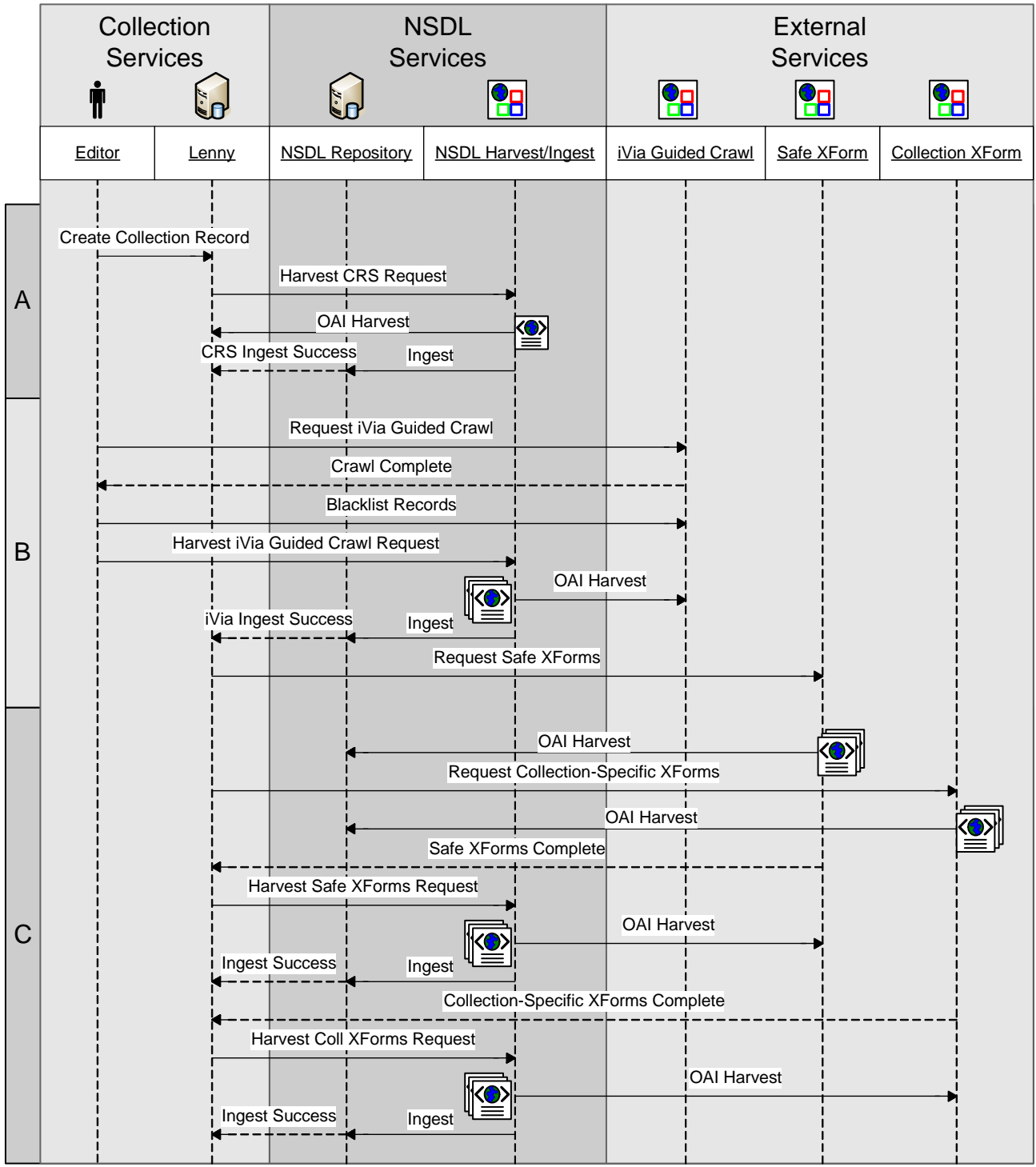

**Figure 1** illustrates a typical sequence of events for a freshly minted collection. Many of the early interactions are actually controlled by an NSDL editor relying on Lenny to communicate with services such as the NSDL OAI-PMH Harvester and the NSDL Repository, that interact with each other to exchange data in the proper sequence (A). The editor also interacts initially with INFOMINE's iVia Expert-Guided Crawler that acts as the primary metadata provider in this scenario (B). Once the primary metadata has been successfully harvested and ingested, Lenny takes over and begins cueing the services playing in this particular orchestration, in this case the Safe and Collection-Specific Transformation Services (C).

## 3. DEVELOPING DIGITAL LIBRARY SERVICES

Thus far, we have described the services from the perspective of Lenny, the conductor. This perspective shows the motivation for each service, and illustrates how services work together to improve the library as a whole, but glosses over the complex parts played by the service providers. In this section, we examine the service provision model from the perspective of the service provider, by examining a particular example, the INFOMINE Project.

INFOMINE is a virtual library of scholarly Internet resources, accessible through a public interface at http://infomine.ucr.edu [16]. It is powered by the iVia Virtual Library Software, an open-source project distributed by INFOMINE under the terms of the GNU General Public License [17], which provides a wide range of automatic collection development tools, many of which have been adapted to provide services for the NSDL and other collaborators.

Lenny—and other external agents—can issue instructions to iVia though the *Remote iVia Service Interface* (RiSI). The RiSI provides a set of services that can be requested by invoking CGI scripts over HTTP. The interface also provides access to feedback about the progress and results of requested tasks, and provides facilities for task logging, completion notification, and result analysis.

### 3.1 Metadata generation services

INFOMINE's simplest role is as a primary metadata generation service for its own collections. INFOMINE is registered with the NSDL as a metadata provider, and a subset of INFOMINE's expert-created records are harvested from its OAI-PMH server by the NSDL's harvest/ingest system (RiSI interaction is not required). Lenny directs the NSDL to ingest the records incrementally, on a monthly cycle, and incorporates them into the NSDL library alongside the item-level records from other providers.

A more complex case arises when INFOMINE generates primary metadata for other collections. For example, suppose a collection holder has selected a variety of high-quality resources, and made them available on a Web site, but not provided descriptive metadata records. In this case, an NSDL editor can nominate INFOMINE as a source of item-level records, and Lenny will use iVia to generate the primary metadata for the collection.

Lenny uses the RiSI to instruct the *iVia Expert-Guided Crawler Service* to generate the necessary item-level metadata, and passes it the collection URL, notification instructions, and a harvest tag. Upon receiving the instruction, iVia launches a new process. It performs an automatic crawl of the website (starting at the provided URL) to discover all the significant resources on the website, then automatically builds item-level metadata records describing each resource. All item-level records are associated with the harvest tag provided by Lenny, and are available for OAI-PMH harvest using that tag for identification. When the process is complete, iVia notifies Lenny, so that Lenny can initiate an OAI-PMH harvest of the metadata records into the NSDL repository.

#### 3.1.1 Metadata augmentation service

The iVia RiSI interface can also provide a metadata augmentation service. Some collection providers have item level metadata available through OAI-PMH, but only supply a few elements. At the minimum, some metadata providers offer only an identifier and a title.

Lenny can augment this item level metadata with much richer descriptions by using RiSI to instruct the *iVia Enhance Metadata Service* to augment a given set of metadata items. When iVia is invoked, it starts a new process by harvesting the specified collection of metadata records from the NSDL repository via OAI-PMH. For each record harvested, iVia extracts the resource URL, then downloads the resource, and assigns it new metadata with its automatic metadata assignment tools, building a new metadata record describing the resource. These new item-level records can then be ingested back into the NSDL repository.

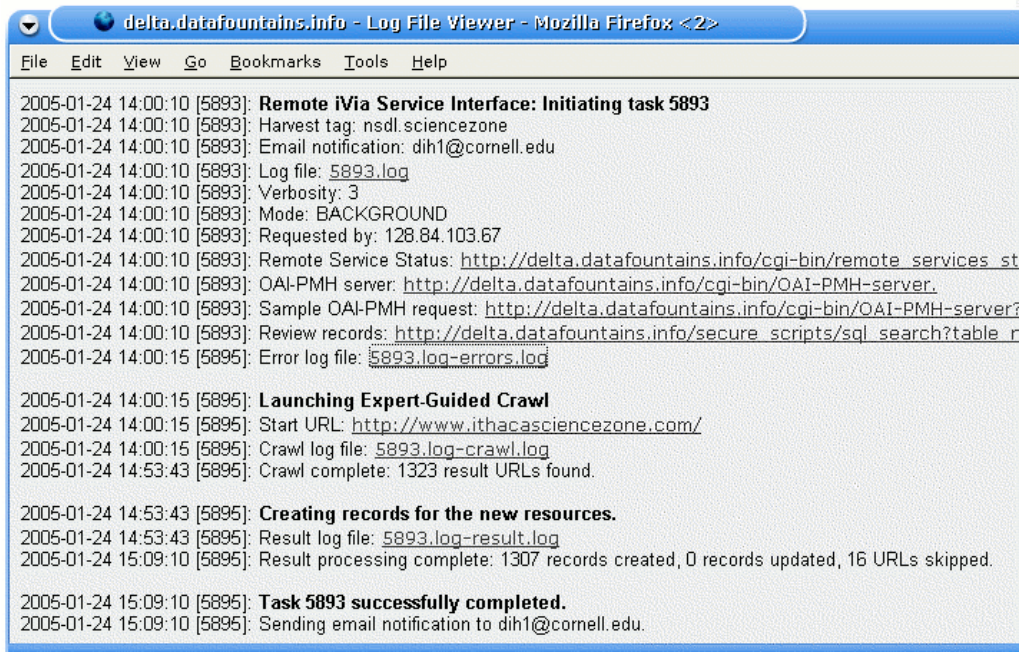

**Figure 2. Expert-Guided Crawler Service task log**

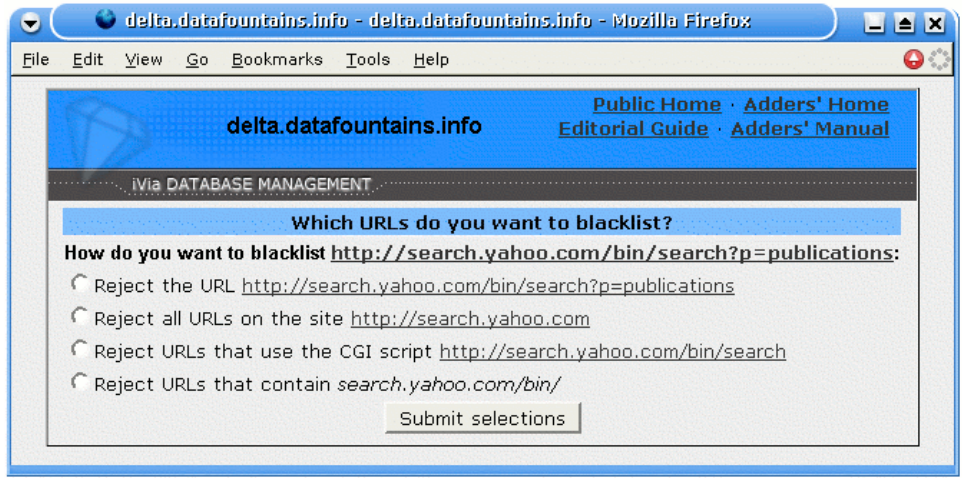

**Figure 3. Expert-Guided Crawler URL blacklisting tool**

When the entire collection is processed, iVia notifies Lenny that the task is complete. When Lenny receives the notification, it directs the NSDL repository to harvest the new metadata, identified as a particular OAI-PMH set, and augment the existing item metadata records available via the NSDL repository's OAI-PMH server.

### 3.1.2 Human Intervention to improve Metadata Services

Many of the tasks that Lenny assigns to iVia involve situations that can best be resolved by expert intervention, so iVia provides several opportunities for human experts to observe and correct its operations.

### 3.1.3 Logging

Every RiSI request is logged in a unique "task log", and possibly other supporting log files. The logs are accessible on the Web using *the iVia Log File Viewer* interface, which reformats logs for readability and inserts links to supporting information, such as iVia records and external Web sites.

Figure 2 shows a task log generated by the Expert-Guided Crawler Service. The task log shows the overall progress of the task, with links to three supporting log files: the crawl log that traces the path of the crawler through the Web site and shows what resources are found, the result log that shows the records that were created for each resource, and the error log that records any resulting URLs for which no records could be created. The error log is particularly useful, as it is used to identify resources that appear in the collection but are no longer active or accessible.

Whenever a RiSI process is completed, a notification email is sent to a nominated NSDL editor containing the URL to view the appropriate log file, thus enabling asynchronous human review.

### 3.1.4 Expert review

Most iVia services do not provide output directly; instead, they build a set of item-level records in the iVia database which can be harvested via OAI-PMH. These result sets can be reviewed over the Web through the *Review Results* interface. Both the task log and the email notification provide a link that directs the expert to the appropriate Web page.

The result list displays each of the item-level records in the specified harvest set, identified by Title and URL. Alongside each item are a set of buttons that let the user view a record, edit, delete it, or blacklist it (see below). The result set can be sorted in different ways.

### 3.1.5 Blacklisting

In some cases, the crawler traverses sections of a Web site that should be avoided, or creates records describing resources that are not suitable for inclusion in the NSDL, such as URLs whose registration has lapsed, links to funding providers or staff members, or mirror URLs. One option for dealing with these problems is to blacklist the offending URLs.

The URL blacklist is a powerful mechanism that excludes a URL (or set of URLs) from future consideration by any automated iVia service. A link to the URL blacklisting tool appears alongside every result in the review results list. Figure 3 shows the blacklisting tool itself. The tool is invoked on a specific URL, and prompts the user to choose a "pattern" to ignore, which is added to the blacklist. Any metadata records whose URLs match a blacklisted pattern are removed from the iVia database, and will not be considered in future crawls. Blacklisting decisions are remembered, and apply to the current process and all future processes.

## 4. CONCLUSION

The NSDL has moved from a traditional collection/item framework to a new, service-oriented model, spurred by the emergence of collections that play a selection role but often do not provide appropriate metadata. This model of service provision holds much potential in an environment where persistent metadata quality issues threaten to overwhelm aggregators hoping to build services on top of harvested metadata. No single aggregator can fill in the quality gaps alone, but if metadata services are built to interoperate with a variety of aggregators using low barrier protocols like

OAI-PMH, many can benefit from the work, freeing resources for new service development.

The INFOMINE services are serving as a model for the development of a diverse set of metadata services, potentially available to other aggregation services in addition to NSDL. These services can be built by a variety of service entities, based on need and available expertise. The growth of these services mirrors to some extent the collaboration and sharing infrastructure prevalent in traditional libraries, although without the supportive leadership of "bibliographic utilities" to serve as central repositories of data. The infrastructure built upon OAI-PMH is thus far more distributed, and less dependent on formal agreements and overt economic considerations. Despite the distribution of effort and lack of formal leadership, there are signs of increasing maturity in the metadata aggregator world.

The service model described here supports a set of collaboratively developed services that can assist a variety of aggregators, as well as resource selection services looking for assistance in developing item-level metadata for their own use.

## 5. ACKNOWLEDGMENTS

The authors would like to thank Professor William Y. Arms and Naomi Dushay of the NSDL team at Cornell for their review of this paper and excellent suggestions for improvement.

## 6. REFERENCES

[1] Dushay, N. and Hillmann, D. I. Analyzing Metadata for Effective Use and Re-use. *Paper presented at the DC2003 Conference* (Seattle, WA, Oct. 2003) http://dc2003.ischool.washington.edu/Archive-03/03dushay.pdf

[2] Shreeves, S. L., Kaczmarek, J. S., and W. Cole, T. W. Harvesting Cultural Heritage Metadata Using the OAI Protocol. *Library Hi Tech,* 21, 2 (2003), 159-169

[3] Guy, M., Powell, A., and Day, M. Improving the Quality of Metadata in ePrint Archives. *Ariadne,* 38, (Jan. 2004). http://www.ariadne.ac.uk/issue38/guy/

[4] Barton, J., Currier, S., and Hey, J. M. N. Building Quality Assurance into Metadata Creation: an Analysis based on the Learning Objects and e-Prints Communities of Practice. *Paper presented at the DC-2003 Conference* (Seattle, WA, Oct. 2003). http://www.siderean.com/dc2003/201_paper60.pdf

[5] Zia, L., et al. The NSF National Science, Technology, Engineering, and Mathematics Education Digital Library (NSDL) Program. *D-Lib Magazine*, 10, 3 (Mar. 2004). http://www.dlib.org/dlib/march04/zia/03zia.html

[6] Lagoze, C., et al. Core Services in the Architecture of the National Science Digital Library (NSDL). *JCDL 2002*. (Portland, OR, 2002). http://arxiv.org/ftp/cs/papers/0201/0201025.pdf

[7] Arms, W. A., et al. A Spectrum of Interoperability: The Site for Science Prototype for the NSDL. *D-Lib Magazine*, 8, 1 (Jan. 2002). http://www.dlib.org/dlib/january02/arms/01arms.html

[8] Hillmann, D. I., Dushay, N. and Phipps, J. *Improving Metadata Quality: Augmentation and Recombination. Paper presented at the DC2004 Conference* (Shanghai, China, Oct. 2004). http://metamanagement.comm.nsdl.org/Metadata_Augmentation--DC2004.html

[9] NSDL Metadata Primer: NSDL Safe Transforms. http://metamanagement.comm.nsdlib.org/safeXform.html

[10] Godby, C. J., Young, J. A., and Childress, E. A Repository of Metadata Crosswalks, *D-Lib Magazine*, 10, 12 (Dec. 2004). http://www.dlib.org/dlib/december04/godby/12godby.html

[11] The DCMI Type Vocabulary. http://dublincore.org/documents/dcmi-terms/#H5

[12] Fielding, R. T. *Architectural Styles and the Design of Network-based Software Architectures,* Ph.D. Dissertation, University of California, Irvine, 2000. http://www.ics.uci.edu/~fielding/pubs/dissertation/top.htm

[13] Common Gateway Interface (CGI) scripts, http://hoohoo.ncsa.uiuc.edu/cgi/

[14] XML-RPC Home page, http://www.xmlrpc.com/

[15] SOAP Version 1.2 Part 0: Primer, http://www.w3.org/TR/soap12-part0/

[16] Mitchell S., Mooney M., Mason J., Paynter G. W., Ruscheinski J., Kedzierski A., and Humphreys, K. iVia Open Source Virtual Library System. *D-Lib Magazine,* 9, 1 (Jan. 2003). http://www.dlib.org/dlib/january03/mitchell/01mitchell.html

[17] GNU General Public License, http://www.gnu.org/copyleft/gpl.html